\documentclass[sigconf]{acmart}

\usepackage{booktabs} 
\usepackage{algorithm}
\usepackage{algorithmic} 

\usepackage{amsmath}
\usepackage{pict2e} 
\usepackage{keyval} 
\usepackage{fp} 
\usepackage{diagbox}

\usepackage{multirow}

\usepackage{graphicx}
\begin{document}
\title{Shared Bottleneck Detecction Based on Trend Line Regression for Multipath Transmission}
\titlenote{Produces the permission block, and
  copyright information}
\author{Songyang Zhang}
\affiliation{
 \institution{School of Computer Science and Engineering, Northeastern University, China}
}
\email{sonyang.chang@foxmail.com}
\author{Weimin Lei}
\affiliation{
  \institution{School of Computer Science and Engineering, Northeastern University, China}
}
\email{leiweimin@ise.neu.edu.cn}
\author{Wei Zhang}
\affiliation{
  \institution{School of Computer Science and Engineering, Northeastern University, China}
}
\email{zhangwei1@mail.neu.edu.cn}
\email{leiweimin@ise.neu.edu.cn}
\author{Yunchong Guan}
\affiliation{
  \institution{School of Computer Science and Engineering, Northeastern University, China}
}
\email{y.c.guan@foxmail.com}

\begin{abstract}
The current deployed multipath congestion control algorithms couple all the subflows together to avoid bandwidth occupation aggressiveness if the subflows of multipath transmission protocol share common bottleneck with single path TCP. The coupled congestion control algorithms can guarantee well fairness property in common bottleneck but result in rate increase conservativeness in none-sharing bottleneck situation. Thus, the throughput of multipath session can be further improved when combing with effective shared bottleneck detection mechanism.
This paper proposes a delay trend line regression method to detect if flows share common bottleneck. Deduced from TCP fluid model, the packet round trip delay signal shows linear increase property during the queue building up process of the narrowest link and the delay trend line slopes of two flows are in close proximity if they traverse the same bottleneck link. The proposed method is implemented on multipath QUIC golang codebase and extensive simulations are performed to validate its effectiveness in detecting out flows traversing common bottleneck. If the subflows are detected out via a common bottleneck, the sender would perform coupled congestion control algorithm and perform congestion control seperately on flow level in none sharing bottleneck case. Results show a multipath session with two subflows can obtain 74\% gain on average in throughput compared with single path connection when Linked Increases Algorithm (LIA)  is in combination with trend line regession shared bottlenck detection algorithm in none shared bottleneck, and show well fairness property in common bottleneck scenarios. 
\end{abstract}


\maketitle

\section{Introduction}
Even though the internet infrastructure has made fast advance in recent years, its most important component Transmission Control Protocol(TCP) basically routes packets through a single path connection, which makes it suffer from several drawbacks. The single path TCP cannot provide continuous connectivity, once the transit networks get stuck in outrage which is not rare event\cite{Holterbach2017SWIFT}. The bandwidth of a single connection may not meet the requirement of some bandwidth consuming applications, for example interactive cloud video games and conference. 

Modern devices equipped with multi-homed interfaces make the concurrent access to heterogeneous networks possible. The datacenter network provides redundant paths for high availability concern and mobile devices can transmit data through different wireless technology (4G and WiFi). In such situation, a connection can split its traffic on different paths with the multipath transport protocol such as CMT-SCTP\cite{Iyengar2006Concurrent} or MPTCP\cite{RFC6182}, which brings out many advantages such as providing aggregating bandwidth, accelerating flow completion time and improving robustness in case of single path route failure.

For its performance improvement, the multipath transmission protocol has been thoroughly researched. One key research point is the multipath congestion control algorithm, which has achieved several fruitful results. One of the operation goals of MPTCP is that a multipath flow should not take up more capacity from any of the resources shared by its different paths, than if it was a single flow using only one of these paths \cite{rfc6356}. Performing congestion control independently on per-path level would result unfair bandwidth occupation if the subflows of a multipath transmission session share the same bottleneck link with single path TCP connection. Due to the lack of effective route topology informationmost, current multipath congestion control algorithms such as wVegas\cite{Cao2012Delay}, LIA \cite{rfc6356}, OLIA \cite{Khalili2013MPTCP}, Balia \cite{peng2016multipath} take coupled congetion control mechanism on connection level. The coupled congestion control algorithms guarantee well protocol fairness to single path TCP in case of its subflows sharing the common bottleneck but result in conservative rate increase in none common bottleneck situation. 

One natural optimization is to combine multipath congestion control algorithm with shared bottleck detection mechanism. The multipath endpoint performs independently congestion control on separate path to improve throughput in none shared bottleneck situation, and falls back to coupled congestion control on shared bottleneck case to guarantee inter-protocol fairness property. 

We further point out here that nearly all the current implemented congestion control algorithms for MPTCP are confined to the window based traffic control framework. But the newly proposed TCP BBR\cite{Cardwell2017BBR} shows excellent performance which has higher link utilization, lower link queue occupation and smaller round trip delay, which makes the current MPTCP not show any advantage in comparison. And we found in experiment, when the multipath session competes bandwidth with TCP BBR to transmit file of same size, its flow completion time is even longer than a single TCP flow. Such performance of BBR puts MPTCP in a dilemma situation. Its goal to improve throughput to motivate deployment thus is in greate challenge. Even though there is advice to upgrade the MPTCP congestion control algorithm to TCP BBR, the implementation is not seen until present. The MPTCP with  BBR congestion control at flow level will break out bandwidth fairness principle in shared bottleneck link situation. The combination of BBR and shared bottleneck detection algorithm may be a solution to avoid such awkward situation, which could be our future focus.

The congestion control law proposed by Jocobson\cite{jacobson1988congestion} is to regulate TCP sending rate according to additive increase and multiplicative decrease(AIMD). At congestion avoidance phase, a tcp session increases its congestion window by $\frac{1}{cwnd}$ if previous sent packet gets acknowledged by its peer. The loss based congestion control algorithms tend filling the queue buffer resource in bottleneck links during the additive increase phase. When the link queue occupancy has exceeded certain threshold, the active queue management mechanisms such as Drop-Tail, RED or newly crafted algorithms CoDel \cite{Nichols2012Controlling} in router would drop incoming packets to ensure the stability and availability of network links. When a packet loss event is detected, the endpoint multiplicatively reduces its congestion window size by half to alleviate congestion or yields bandwidth for newly coming flows.

Once the total sending rates of end users beyond the capacity of narrow link, the extra sending packets would be queued and the end to end delay would increase. In AIMD situation, such delay increase process shows linear feature and the flows traverse the same bottleneck would observe similar delay accumulated trendline. 

Based on this observation, a delay trendline regression (TR) method is proposed here for shared bottleneck detection. The delay increase slopes in close proximity indicate that the subflows via a common bottleneck in great possibility. The method is theorized from TCP fluid model and the full proof is provided later. The algorithm only relies on end to end delay measurement and no other extra active probe packets are needed here. And the method is implemented on multipath QUIC protocol and simulations show that it can improve bandwidth fairness when multipath TCP session coexists with single path TCP session in shared bottleneck situation and improve its throughput in none shared bottleneck case.

The rest of this paper is organized as follows. Section 2 provides a brief review of related works on shared bottleneck detection technology. The conservativeness of the multipath coupled congestion control algorithm is analysed in Section 3. Our proposed mechanism is described in detail in Section 4, and mathematical analysis based on TCP fluid model is provided to prove the reasonableness of this method. The simulation results are presented on Section 5. Section 6 is the conclusion and some ideas on future research.
\section{Related work}
It's a long time since researchers intuited that the flows traversing a common bottleneck show some similarities. But no validated theoretical proof is provided. And the signals can be used to detect out such similarities is quite limited. The signals that can be exploited for such purpose including inter packet arriving delay, round trip time, one way delay and packet loss rate. Hence, the research results on shared bottleneck detection area are no quite fruitful. 

The ability to detect shared bottleneck for flows has several practical applications. 
The work \cite{Balakrishnan1999Integrated} explored to perform congestion control over aggregated flows. In a typical video chat application, two end user may initiate concurrent flows for different purpose such as file transfer, RTP-base video streaming traffic and voice traffic. The capacity competition of these flows would increase packet loss and link delay if they flow through the same bottleneck. It will make the real time traffic suffer from low quality. A web client would create multiple TCP connections to simultaneously download content from a web server to increase download speed. The user tends to create as many tcp connections as possible from egoistic perspective if the download speed proportional to the number of tcp connection is assumed, which results in vicious competition and leaves the routing path in unstable status. To aggregate these flows into one single virtual flow and perform aggregated congestion control would stabilize the internet and promote fair bandwidth occupation. Overlay network \cite{Andersen2002Resilient} has been proposed as a mean to improve performance, in which the application can exploit parallel paths \cite{Zhang2014general,Guan2018Scalable} to aggregate bandwidth to provide higher quality of service or backup path to resist connection disruptions. One key point is to detect whether the overlay paths share congestion point at the undelay network. The sharing bottleneck in the underlay network can impact the effectiveness of the overlay network.

In \cite{Katabi2001passive}, An entropy based method is proposed for flow clustering, which takes inputs of two consecutive packets inter arriving delay. The rationale behind is that the bottleneck link most of the time is busy processing packets and the inter-packet leaving space depends on bottleneck capacity. The packets that  show the minimum randomness and the lowest entropy in term of inter-packet delay can be grouped together. It should be pointed out the observer should work at the intermediate router and must get all the flows inter-packet space from the upstream. This technology can hardly be applied by end users.

Rubenstein et al \cite{Rubenstein2002Detecting} proposed delay correlation and loss correlation to decide if two flows have shared points of congestion based on end to end measurement. The authors indicates that packets of different flows passing through the same congestion link would have strong correlation in delay and loss. Poisson packets in which the inter packets departure time  follows the Poisson process are injected into links for information collection. When the measurement of correlation between flows is larger than the measurement of correlation within a flow, it's concluded that two flows share points of congestion. Most of the works hereafter inherited this delay correlation method.

The work of \cite{Younis2005FlowMate} adopted Rubenstein’s delay correlation in a busy server for flows clustering with passive delay measurement. The delay information is piggybacked on feedback packets, thus the header of  TCP is extended for such purpose. It was applied on coordinated congestion management to improve resource allocation fairness and network responsiveness. Cui et al \cite{Cui2004SCONE} proposed a tool named SCONE to estimate sharing congestion among internet paths. It is based on correlated packet drops. If the time interval between two packet loss events from different connections is under a predefined threshold, connections are concluded passing the same congestion path. In \cite{Karacali2008Network}, the inter packet arriving jitter is used to infer path intersections at the receiver side. Probe stream and signal stream are injected inject into paths separately. The back to back packets in signal stream would cause the packets in probe stream delayed or lost if the two paths are intersected, and the jitter can be noticed at receiver. This method requires the communication and coordination among nodes to start the test at approximately the same time for jitter signal matching. And it is allergic to the noise from background traffic. In \cite{Wang2004Passive} the authors proposed TCP rate correlation to detect path sharing between paths. The rate change in TCP flows can be seen as a reflection to network conditions considering the TCP bandwidth fairness property.
	
There are also some works apply signal processing technology on shared bottleneck detection. In \cite{Zhu2013TCP}, Fourier transform with the inter packet arriving delay signal is used to detect if  TCP sessions sharing a common path. The work of \cite{Kim2008Wavelet} applied one way delay correlation  with wavelet denoising. The wavelet transform technology is applied to enhance the robustness of the correlation computation to filter out delay variation, in consideration of that the assumption the queue delay on none bottleneck link is close to zero may not hold true in real internet. At first, we prepared to apply Fast Fourier Transform Algorithm on sharing bottleneck detection. During the experiment, we found most of the key frequencies are at the range of (0,1], which makes the comparison of the key frequencies quite hard considering the error impact. Therefore, the round trip delay signal collected by the passive method without active probe packets injecting into network does not show obvious periodic characteristics. In the end, such idea was abandoned.

Hayes et al. \cite{Hayes2014Practical} use summary statistics similarity characteristics (SSC) with parameters of skewness, variation and key frequency to group flows. This method is further implemented on MPTCP for throughput improvement in \cite{Ferlin2016Revisiting}. We implement SSC algorithm on multipath QUIC and performance is compared between SSC and TR.

\section{Throughput analysis of MPTCP}
The throughput of TCP at equilibrium can be deduced from fluid model. On every received acknowledgement, the sender congestion window increment is denoted by I, when packet loss event happened, the window decrease is denoted by D.
According to \cite{peng2016multipath}, the differential equation is:
\begin{equation}
\label{eq1}
\dot x=\frac{x}{rtt}(I(1-p)-Dp)
\end{equation}
for TCP Reno, $I=\frac{1}{w}, D=\frac{w}{2}$. The equilibrium rate will get when $\dot x=0$, and the average throughtput of TCP Reno is $\frac{1}{rtt}\sqrt{\frac{2(1-p)}{p}}$.

Since most of the packets flowing through internet are transported by TCP, the protocol friendliness of proposed congestion control algorithms is a main concern. The friendliness of multipath congestion control mechanism at the bottleneck is an important property. Current proposed methods mainly take the coupled congestion control to avoid bandwidth aggressiveness in competing with single path TCP. The coupled window control law of LIA is $I_r=\frac{\max(w_k/{rtt_k}^2)}{\{\sum_{k\in s}w_k/rtt_k\}^2}, D_r=\frac{w_r}{2}$. 

Substituting $I_r, D_r$ into equation \eqref{eq1} and let $\dot x=0$, the sum of average throughput on all sub-paths of MPTCP can thus be obtained:
\begin{equation}
\label{eq2}
\sum_{k\in s} x_k=\max_{k\in s}\frac{1}{rtt_k}\sqrt{\frac{2(1-p_k)}{p_k}}
\end{equation}

From the throughput formula \eqref{eq2}, the coupled MPTCP makes the aggregate throughput should not exceed the throughput of a single flow using only the best path. This congestion window control mechanism guarantees well fairness property when the subflows of MPTCP transport through the same bottleneck and compete resource with single path TCP. But the rate control of MPTCP becomes too conservative in none sharing bottleneck situation. In this case, if the each subflow of MPTCP behaves like a single path TCP, its throughput on all path in theory could reach:
\begin{equation}
\label{eqpathsum}
\sum_{k\in s} x_k=\sum_{k\in s}\frac{1}{rtt_k}\sqrt{\frac{2(1-p_k)}{p_k}}. 
\end{equation}

The average throughput of a MPTCP session regulated from flow level congestion control is greater than its coupled form congestion control. Thus, the throughput of multihomed endpoint would be further improved if the combination of the shared bottleneck detection algorithm and congestion control would be exploited.
\section{Shared bottleneck detection}
Our method mainly is based on the observation that the delay increase shows linear feature during link queue filling process. If two flows traverse the same bottleneck, the delay increase slope of two flows will be in close proximity during link queue building up phase, which can be proved from TCP fluid model. Some extra assumptions should be put forward here for subsequent theoretical deduction.

First, the network is dominated by responsive flows and a responsive flow would react to congestion signal by rate reduction. The assumption holds true based on the fact that most flows in real network are transmitted via TCP protocol and the RTP-based streaming traffic has deployed a variant of AIMD congestion control algorithm in WebRTC. Due to lack of a centralized network resource allocation operator, the function of network congestion control is moved to endpoint and bandwidth is probed in a distributed way. The endpoint would increase its sending rate for bandwidth probe. This process tends to increase end to end delay when the total rates of all flows exceed the narrow link capacity. When the link buffer in full state, the active queue management mechanism deployed in router will drop the following incoming packets. The packets loss can be seen as link congestion signal. The network congestion signal must be detected by certain flows. Network congestion is not caused by a single flow and the congestion avoidance action of a several flows obviously cannot alleviate network congestion status. Only when certain number of flows get the signal and react it by decreasing rate, the network congestion can be alleviated. The rate decrease action of endpoints would let the congestion link to drain queue buffer accumulated before. 

Second, the packet queue delay is mostly depended on bottleneck. The end to end delay of a packet is composed of transmission delay and queue delay. When a packet  traverses non bottleneck links, the intermediate routers have empty queue with large possibility and the packet will be transmitted out immediately. As the end users increase rate to probe more available bandwidth, the packets will first queue at the bottleneck link when the total rates of all users exceed the narrow link capacity. 

Such assumptions make sure that the delay trendline slope during link queue increase process can be a reflection of internal property of bottleneck link. The delay signal noise introduced from unresponsive background traffic and extra packet queue delay in none bottleneck can be neglected for theoretical analysis convenience.
\subsection{Mathmatical model and analysis}
Let there be n flows through a bottleneck link with capacity $C$ and a flow rate of user $i$ is denoted by $x_i$. $Q$ denotes the bottleneck queue occupation length. The multifluid model in \cite{Hong2003Note} is followed here. According to the AIMD control law, in a time unit $\Delta t$, the number of acknowledgement packets received by the sender $i$ is $a_i(t)\Delta t$, where $a_i(t)$ is the packets acknowledgement rate. And the congestion window increase during this time unit is $\Delta w_i(t)=\frac{a_i(t)\Delta t}{w_i(t)}$.
\begin{equation}
\label{eqwindow}
\dot w_i(t)=\frac{\Delta w_i(t)}{\Delta t}=\frac{a_i(t)}{w_i(t)}
\end{equation}

When the total rates of all network users does not exceed the bottleneck capacity, the received packet at the Intermediate node could be sent out immediately and no packet would be buffered. Once the total rates exceed the bottleneck capacity, the extra sent packets would be buffered. The queue length that increases over time can be denoted by $Q=\int{(\sum_{i=1}^{n}x_i-C)}\mathrm{d}t$. The queue length variation can be expressed by the following equation. 
\begin{equation}
\label{eqqueue}
\dot Q(t)=\left\{
\begin{array}{rcl}
0,&& {\sum_{i=1}^{n}x_i\leq C}\\
\sum_{i=1}^{n}x_i-C.&& {\sum_{i=1}^{n}x_i>C}
\end{array} \right.
\end{equation}

On every rtt, one more packet would be sent out, hence the sending rate of an endpoint during the link queue increase phase can be denoted by:
\begin{equation}
\label{eqrate}
x_i(t)=\frac{a_i(t)r_i(t)+1}{r_i(t)}
\end{equation}

When the sending rates of all users exceed the capacity of a bottleneck, the packets will be queued and spaced by the bottleneck. And the total acknowledgement rates are exactly equal to link capacity, which can be mathematically expressed as:
\begin{equation}
\label{eqsum}
\sum_{i=1}^{n}a_i(t)=C 
\end{equation}

Combining equation \eqref{eqqueue}, \eqref{eqrate} and \eqref{eqsum}:
\begin{equation}
\label{queuedelay}
\dot Q(t)=\sum_{i=1}^{n}[a_i(t)+\frac{1}{r_i(t)}]-C=\sum_{i=1}^{n}\frac{1}{r_i(t)}
\end{equation}

From the perspective of an end user, the round trip delay a packet experienced was mainly comprised by propagation delay and link queue delay. $r_{min}$ is the end to end propagation delay.
\begin{equation}
\label{eqdelay}
r_i(t)=r_{min}+\frac{Q(t)}{C}
\end{equation}

The derivative of r is:
\begin{equation}
\label{rttderivative}
\dot r_i(t)=\frac{\dot Q(t)}{C}
\end{equation}

One of key feature of TCP congestion control is its fairness\cite{Chiu1989Analysis}, which guarantees all the flows through the same bottleneck link have similar throughput at equilibrium, $x_i\approx x_j$. All the flows behave in homogeneity is thus obtained, the equation \eqref{queuedelay} can be rewritten:
\begin{equation}
\label{queuerewrite}
\dot Q(t)=\frac{n}{r_i(t)}
\end{equation}

Combining equations \eqref{rttderivative} and \eqref{queuerewrite}:
\begin{equation}
\label{rttresult}
\dot r_i(t)=\frac{n}{Cr_i(t)}
\end{equation}

Solving the above differential equation:
\begin{equation}
\label{delay-law}
r_i(t)=\sqrt{\frac{2nt}{C}+r_{min}^2}
\end{equation}

The coefficient $\frac{C}{n}$ can be considered as a fair bandwidth share of a TCP connection. The path transmission delay is a small value in unit of milliseconds in real notwork. And $\frac{n}{C{r_{min}}^2}$ is also considered as a small value. The equation \eqref{delay-law} can be approximated by:
\begin{equation}
\label{delay-law-approximate}
r_i(t)=\frac{nt}{C}+r_{min}
\end{equation}

The maximum queue length of the link is denoted by $Q_{max}$. When the queue buffer overflows, packets loss event happens and flows will back off its rates to let queue drain. The variation range of $r_i(t)$ can be gotten: $r\in [r_{min},r_{min}+\frac{Q_{max}}{C}]$.

According to equation \eqref{delay-law-approximate}, when the total rates exceed the capacity of bottleneck link, the delay trendline slope in equation \eqref{slope} is closely related to the number of flows traversing through it and the capacity of the bottleneck link. In network configuration, different paths may have similar packet handling capacity in its narrow link, but the number of flows through these paths may vary in great possibility. The delay trendline slope could be treated as a reflection of inherent property of bottleneck link. And the different sessions that traversing the same bottleneck link would have similar trend line slope value. Thus, the comparison of the trendline slope value is proposed to decide whether the subflows of the multipath session via the same bottleneck.

\begin{equation}
\label{slope}
slope=\frac{n}{C}
\end{equation}

The above analysis is mainly based on the classic TCP congestion control law. But we argue that the same conclusion can be applied as long as other TCP congestion control variants follow the additive increase principle during the congestion avoidance phase. The most widely deployed TCP congestion control algorithm in Linux kernel is Cubic. Its congestion window increase can be approximated by a linear equation in \cite{Ha2008CUBIC}. And in later part we will show that the method works well in TCP Cubic through simulation.
\subsection{Evidence from simulation and real trace}
\begin{figure}
\includegraphics[height=2.5in, width=3in]{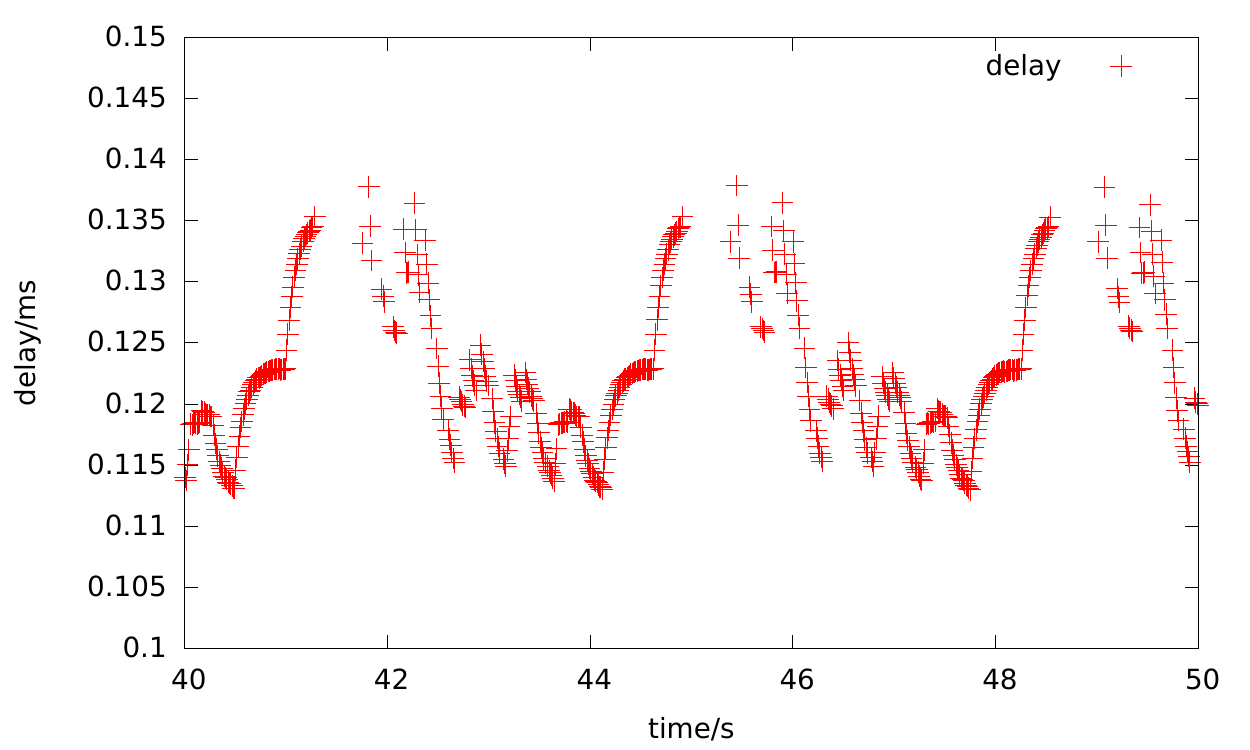}
\caption{The packet round trip delay of single TCP flow}
\label{Fig:p2pdelay}
\end{figure}
\begin{figure}
\includegraphics[height=2.5in, width=3in]{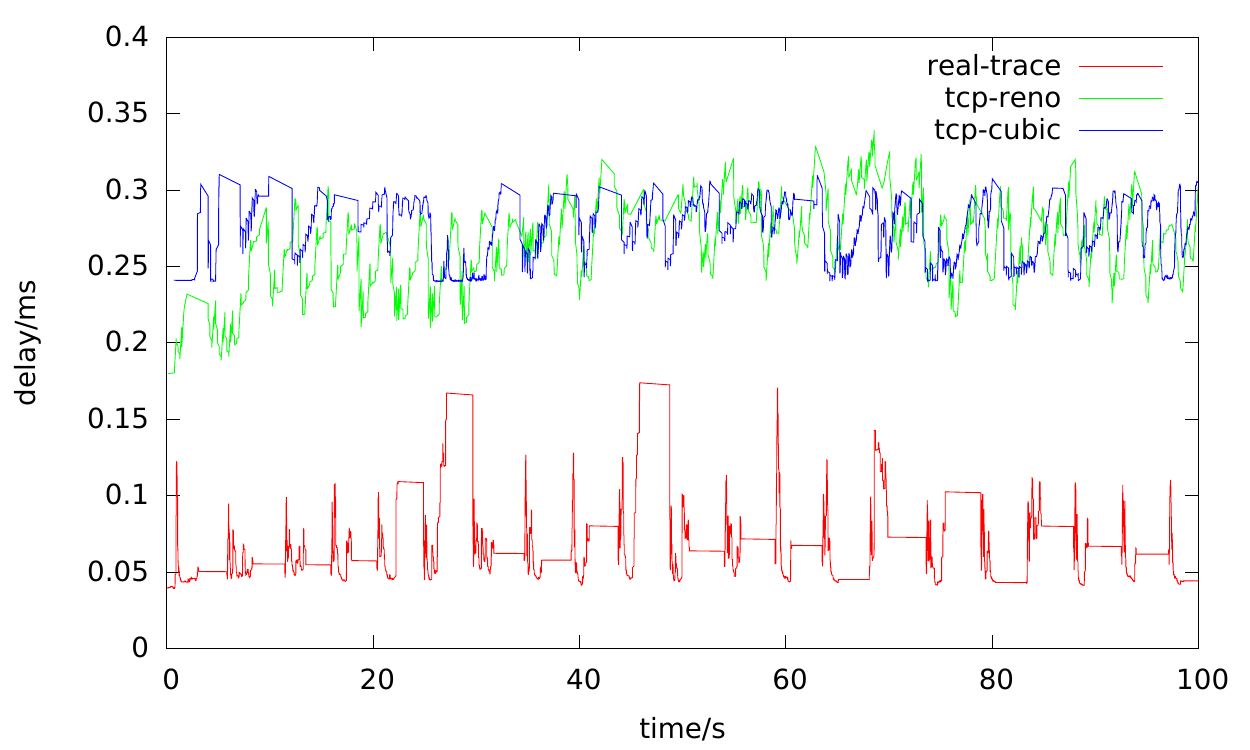}
\caption{The RTT of TCP flows}
\label{Fig:realtrace}
\end{figure}
\begin{figure}
\includegraphics[height=0.5in, width=2.5in]{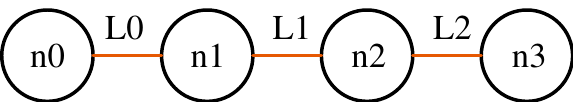}
\caption{Experiment topology}
\label{Fig:n4l3}
\end{figure}
\begin{table} 
\caption{Simulation configuration}
\label{tab:simu-config} 
\scalebox{0.6}{
\begin{tabular}{|l|ccccc|}
\hline
\diagbox{Experiment}{Config}{Parameter} & L0 & L1& L2&flow number&congestion protocol\\
\hline
1 &(500Mbps,20ms)&(20Mbps,20ms)&(500Mbps,50ms)&30&TCP Reno\\
2 &(100Mbps,50ms)&(40Mbps,20ms)&(100Mbps,50ms)&30&TCP Cubic\\
\hline
\end{tabular}}
\end{table}
To validate the deduced conclusion from the fluid model, experiments are conducted both on ns3 \cite{ns3} simulation environment and real internet data trace. 
A point to point topology was created on ns3 with link rate 1Mbps and 50ms one way transmission delay. A single TCP Reno flow was initiated. As showed in Figure \ref{Fig:p2pdelay}, the rtt curve shows root square growth in delay accumulated period, which confirmed the correctness of equation \eqref{delay-law}. 

To further validate the results in multiflow case, a four nodes with three links topology was created shown in Figure \ref{Fig:n4l3} and 30 flows were initiated. The links configuration is shown in Table \ref{tab:simu-config}. In the first simulation case, the minimal rtt is about 180ms the packets generated by all 30 flows are transmitted by TCP Reno. And in the second configuration, the minimal rtt is about 240ms and all the packets are transmitted by TCP Cubic. The RTT delay traces of a flow in each experiment are collected. In order to collected RTT data in real internet, several experiments were conducted to upload a 1G  random generated binary file to Baidu Wangpan \cite{wangpan} from a Windows host. The real-trace curve in Figure \ref{Fig:realtrace} shows the results. And the linear delay increase process is indeed observed from the three RTT curves, which can be explained by the RTT approximate equation \eqref{delay-law-approximate}, no matter what the data sources come from simulation trace or real network trace.
\subsection{Algorithm implementation}
As for the multipath transmission scheme, it’s not rare for two paths with different delay. Even in a single path, a packet can be router through several links and queued in intermediate routers. A packet queued in none bottleneck link will introduce extra noise into the delay signal. In such situation, the traditional correlation technology may not perform well due to the impact of noise and path heterogeneity.

The proposed SBD algorithm is deployed on sender side and the RTT data trace is collected. For every incoming acknowledgement packet, the tuples on the timestamp it arriving and the newly computed round trip delay are collected. It should note that the RTT signal contains much noise, which may introduced by the packet queue at the none bottleneck link and the queued acknowledge packet delay on the reverse link. As we argue before, the delay increase is mostly the result of packet queued at the bottleneck link and several outliers during the link queue filling process will not have great impact on the delay slope computation with the regression method.
\begin{figure}
\includegraphics[height=2.5in, width=3in]{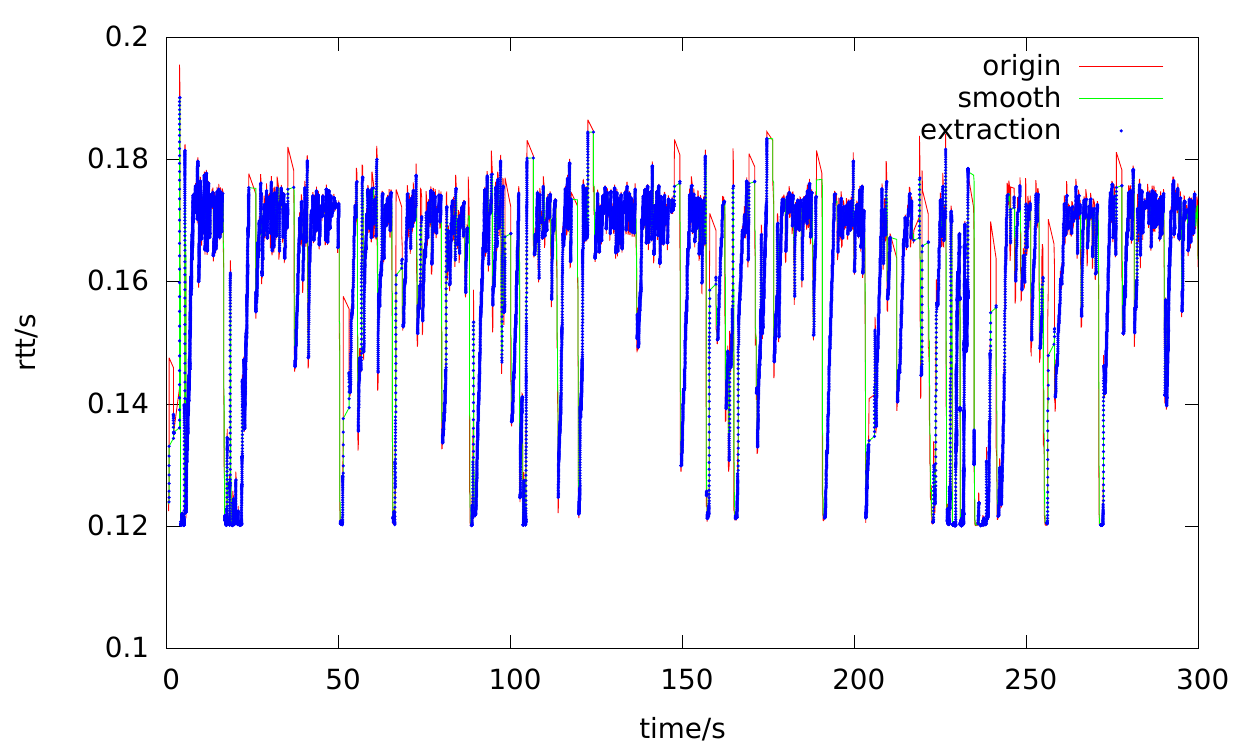}
\caption{Data processing example}
\label{Fig:example}
\end{figure}
\paragraph{(1) Samples extraction during link queue builing up phase}
\ \

The key component of the algorithm is the linear regression of delay signal during the queue accumulation period. How to effectively extract these data points out during such period is none a trivial task, since these data samples during the none congestion period and link queue drain up phase are also collected. An algorithm based on adjacent values comparison is used for data points grouping, which is named as max filter. These data points with the monotonous increasing delay are belonging to the same queue building period and can be grouped together for later slope computation. A data trace of a simulation experiment is shown in Figure \ref{Fig:example}. Even though the deduced theory indicates the delay curve shows linear increase trend in link queue build phase, it can be observed in the original-curve that even in such period, the original RTT curve is not monotonous increasing and the slope curve shows spikes. First the exponential smoothing filter is carried to filter out such spikes. With experiment, $\alpha=0.9$ meets the requirement.
\begin{equation}
\label{smooth-filter}
s(t+1)=\alpha *rtt+(1-\alpha)*s(t)
\end{equation}

\begin{algorithm}[htb] 
\caption{Delay data extraction} 
\label{alg:extration} 
\begin{algorithmic}[1] 
\REQUIRE ~~\\ 
The array of timestamp, $time\_array$;\\
The array of smoothed end to end delay, $delay\_array$;\\
the size of array $N$;
\ENSURE ~~\\ 
A two-dimensional array records the time stamp value of all groups, $group\_x$;\\
A two-dimensional array records the delay value of all groups, $group\_y$;\\
the number of data points of group i, $group\_z[i]$;\\
size of groups, $g$;
\STATE $last=delay\_array[0]$;
\STATE $last\_index\gets 0$;
\STATE $g \gets 0$;
\STATE $counter \gets 0$;
\FOR{each $i \in [1,N)$}
\IF{$delay\_array[i]\geq last$}
\IF{$last\_index==0$}
\STATE $g \gets 0$;
\STATE $counter\gets 0$;
\ELSIF{$i-last\_index>1$}
\STATE $g+=1$;
\STATE $counter \gets 0$;
\ENDIF
\STATE $group\_x[g][counter]=time\_array[i]$;
\STATE $group\_y[g][counter]=delay\_array[i]$;
\STATE $last\_index\gets i$;
\STATE $counter++$;
\STATE $group\_z[g]\gets counter$;
\ENDIF
\STATE $last\gets delay\_array[i]$;
\ENDFOR
\end{algorithmic}
\end{algorithm}

\begin{algorithm}[htb] 
\caption{Groups merge} 
\label{alg:merge} 
\begin{algorithmic}[1] 
\REQUIRE ~~\\ 
$group\_x, group\_y$;\\
the number of data points of group i, $group\_z[i]$;\\
number of groups,$G$;
\ENSURE ~~\\ 
The values of merged groups, $merge\_group\_x,merge\_group\_y$;\\
the number of data points of merged group i, $merge\_group\_z[i]$;\\
size of merged groups, $S$;
\FOR{each $i \in [0,G)$}
\STATE $z \gets group\_z[i]$;
\STATE$average\_group[i]=\frac{\sum_{j=0}^{z-1}group\_y[j]}{z}$;
\ENDFOR
\FOR{each $i \in [0,G-1)$}
\STATE $z \gets group\_z[i]$;
\STATE $gap \gets group\_y[i+1][0]-group\_y[i][z-1]$;
\STATE $\overline {sRTT}\gets average\_group[i+1]$;
\IF{$(groupt\_y[i][z-1]<\overline {sRTT})\&\&(gap<\delta)$}
\STATE $next\_group[i]=i+1$;
\ELSE
\STATE $next\_group[i]=-1$;
\ENDIF
\ENDFOR
\STATE $g\gets 0$;
\STATE $counter\gets 0$;
\FOR{each $i \in [0,G-1)$}
\STATE $z \gets group\_z[i]$;
\FOR{each $j \in [0,z)$}
\STATE $merge\_group\_x[g][j]\gets group\_x[i][j]$;
\STATE $merge\_group\_y[g][j]\gets group\_y[i][j]$;
\STATE $counter++$;
\STATE $merge\_group\_z[g]\gets counter$;
\ENDFOR
\IF{$next\_group[i]==-1$}
\STATE $g++$;
\STATE $counter\gets 0$;
\ENDIF
\ENDFOR
\STATE $S\gets g$;
\end{algorithmic}
\end{algorithm}

After filtering, the data extraction becomes easier as most of the sample points during queue filling phase shows monotonous increasing features. These data samples then are fed back into the max filter algorithm, and a large part data samples that belong to none link congestion phase can be filtered out. This max filter is described in Algorithm \ref{alg:extration}. 

Figure \ref{Fig:example} works as an example. The smoothed-curve is results of the original data after exponential smoothed filter processing. And the data-extraction points curve is the results after the processing of Algorithm \ref{alg:extration}. This curve clearly shows that this algorithm can effectively extract out these data points of the queue building up period.
\paragraph{(2) Groups merge}
\ \

But the above operation would introduce another problem. After the processing of data samples extraction, the data points belong to the same slope may be partitioned into different groups. The reason is that not all the spikes in the delay curve can be filtered out by the operation of exponential smoothing filter. To further reduce the error, the adjacent groups are merged based on the following two defined conditions. The first condition is to decide whether the average smoothed delay value of $group_i$ is larger than the largest smoothed delay value of $group_{i-1}$. Since the delay value of a group is monotonous increasing after smoothed processing, the largest delay value of a group is the coordinate y-axis value of the last point. The second is the time gap between two adjacent groups should less than $\delta$. Here, choosing the appropriate value of $\delta$ is quite tricky and in our experiment is 50 milliseconds. A large value of $\delta$ may lead these data samples belonging two different queue filling process are merged into the same group and introduce large error in slope regression operation, while a smaller value will not work effectively to merge data samplings of a queue filling phase into a group. The group merge processing is described in Algorithm \ref{alg:merge}. 
\begin{equation}
\label{eq:regression}
slope_i=1000*\frac{\sum_{j}\{(x_{ij}-\overline x-x_{i0})*(y_{ij}-\overline y)\}}{\sum_{j}(x_{ij}-\overline x-x_{i0})^2}
\end{equation}
\paragraph{(3) Shared bottleneck decision}
\ \

After these preprocessing operations, the slope of a merged group is thus computed based on linear regression as equation \eqref{eq:regression} shows. $x_{ij}, y_{ij}$ respectively denote the timestamp value and smoothed round trip delay of point $j$ in merged group $i$. Since the unit of the delay value is milliseconds, if this value is directly used in computation, the finally value of slope will be too small. For comparison convenience, the slope value is multiplied by 1000. In other word, the unit of delay slope is $ms/s$

The process to detect shared bottleneck link is mainly based on delay trend line slope comparison. The proposed algorithm is executed in every 5 seconds, and the slope of a group with the most sampling points is used during this time period. This time span of this slope with the most points represents the longest queue filling period. If the other flow traverses the same bottleneck, it will observe a similar delay growth period within a close time proximity. Hence, the conditions to decide whether two sub-flows share the same bottleneck are: the slope comparison error is under a predefined threshold, the time spans of two slopes are overlapped or its time gap is under a threshold $\tau$ in none overlap situation, which is defined 1 second during this experiment. The time gap threshold $\tau$ is mainly a consideration of delay heterogeneity of different paths. The acknowledge packets during the same queue filling process of the bottleneck link will be fed back at different time to the sender in multipath transmission scheme. The comparison error of two slopes belonging to different subflows is defined in Equation \eqref{eq:error}. The predefined error threshold $\epsilon$ is 0.2. The time start point is denoted by $Sn$ and the end time is denoted by $En$ of a merged group of path n. The shared bottleneck decision algorithm is described in Algorithm \ref{alg:sbd}.
\begin{equation}
\label{eq:error}
error=\frac{||slope1-slope2||}{\max(slope1,slope2)}
\end{equation}
\begin{algorithm}[htb] 
\caption{Shared bottleneck decision} 
\label{alg:sbd} 
\begin{algorithmic}[1] 
\REQUIRE ~~\\ 
$slope1, S1, E2, slope2, S2, E2$;
\ENSURE ~~\\ 
Shared bottleneck detection result, $r$;
\STATE $r\gets 0$;
\STATE $s=max(S1,S2)$;
\STATE $e=min(E1,E2)$;
\IF{$e-s<0\ $\AND$||e-s||>\tau$}
\RETURN {$r$}
\ENDIF
\STATE $error\gets\frac{||slope1-slope2||}{\max(slope1,slope2)}$
\IF{$error\leq \epsilon$}
\STATE $r\gets 1$
\ENDIF
\RETURN {$r$}
\end{algorithmic}
\end{algorithm}
\section{Simulation Results}
As pointed by \cite{Hayes2014Practical}, to verify the shared bottleneck algorithm on real network environment has several inconveniences. The information on the network link occupation is hard to get, if not possible. And, what’s important, even the implemented algorithm gives positive result, the lack of detailed network link topology information makes it hard to differentiate between false positive and true positive.

Hence, most of the proposed algorithms assess effectiveness based on simulated topology. The experiments on a simulation environment have several advantages. More network topologies can be built, the parameters of link bandwidth and delay are configurable. And the finally results are little influenced by external uncontrollable interface and can be fully reproducible. Since the work of \cite{Ferlin2016Revisiting} has already shown the effectiveness and performance improvement of the implementation of SBD algorithm on MPTCP on real network testbed, 
the verification and comparison of the proposed idea of this paper is mainly conducted on simulation platform on ns3 and Mininet.

The experiments are divided into two parts. First, the data traces are collected and analyzed on ns3 platform. TCP Reno and Cubic flows are running on different active queue management mechanisms Drop tail and Red separately. Seconds, the proposed algorithm Delay Trend-line Regression was implemented on the code base of multipath QUIC\cite{DeConinck2017Multipath}. 
And the performance comparison with similar statistical characteristics (SSC) in \cite{Ferlin2016Revisiting} in terms of throughput improvement and bandwidth occupation fairness was conducted on Mininet \cite{Lantz2010Network}. The reason to choose multipath QUIC other than MPTCP in Linux kernel is the complexity of MPTCP. The code change and compiling is related to the kernel, which is time consuming and error prone. And Therefore, multipath QUIC  even performs better than MPTCP shown in \cite{DeConinck2017Multipath}.
\subsection{Results from trace analysis}
\begin{figure}
\includegraphics[height=0.8in, width=3in]{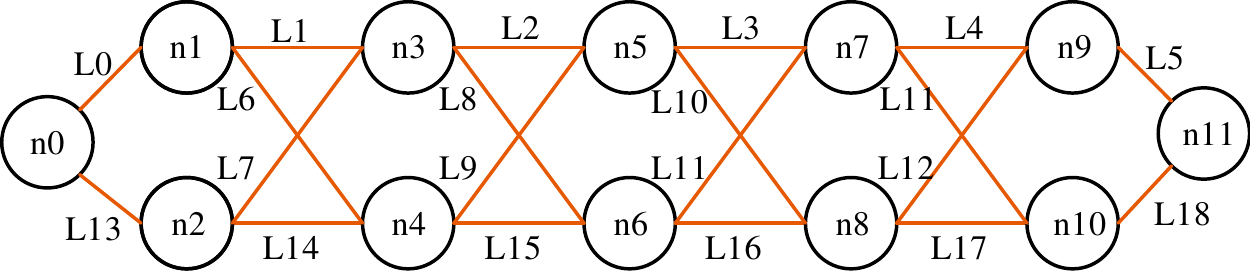}
\caption{Shared bottleneck detection topology}
\label{Fig:topology}
\end{figure}

The topology for shared bottleneck detection in this experiment shows in Figure \ref{Fig:topology}, which could be configured into different sub-topology with different nodes and links, including shared bottleneck and none shared bottleneck situation. 

The effectiveness of proposed algorithm is first validate based on offline trace data analysis. The traced data samples from different TCP connections are collected from ns3 simulation platform. The intermediate nodes act as router are configured into two different queue management algorithms, FIFO (drop tail) and Red. Testing our method in Red queue management is mainly the consideration that the global synchronization phenomenon in Drop tail queue management causes the rate dynamics of all the TCP flows via the same congested link following quite similar behavior. These experiments with RED queue management configuration also work as example to show that our method is not restricted from queue management algorithm. Different TCP congestion control algorithms Reno and Cubic are tested. Considering our method is induced from classical AIMD algorithm and the current Linux kernel default congestion control mechanism is Cubic. To step further, the proposed method is tested in Mininet platform with the newly implementation congestion control TCP BBR and the experimental results show its effectiveness in BBR.  

\begin{table*}[]
\caption{dumbbell topology configuration}
\label{Tab:dumbbell}
\begin{tabular}{|c|c|c|c|c|c|c|c|c|c|c|}
\hline
\multirow{2}{*}{} & L1           & L2           & L3           & L7           & L10           & \multicolumn{3}{c|}{RED} & \multirow{2}{*}{flows(n1-n7)} & \multirow{2}{*}{flows(n2-n8)} \\ \cline{2-9}
                  & \multicolumn{5}{c|}{(B(Mbps), D(ms), Q(ms))}                                             & minTh& maxTh& QL &                               &                               \\ \hline
E1                 & (500, 25, 100) & (40, 20, 100)  & (500, 20, 100) & (500, 50, 100) & (500, 50, 100) & 100       & 120      & 140   & 15                            & 15                            \\ \hline
E2                 & (500, 25, 100) & (60, 20, 100)  & (500, 30, 100) & (500, 50, 100) & (500, 50, 100) & 100       & 120      & 140   & 15                            & 15                            \\ \hline
E3                 & (500, 10, 100) & (90, 20, 100)  & (500, 10, 100) & (500, 20, 100) & (500, 20, 100) & 100       & 140      & 200  & 30                            & 30                            \\ \hline
E4                 & (500, 10, 100) & (120, 20, 100) & (500, 10, 100) & (500, 50, 100) & (500, 50, 100) & 100       & 200     & 220  & 30                            & 30                            \\ \hline
E5                 & (500, 10, 100) & (120, 20, 100) & (500, 10, 100) & (500, 30, 100) & (500, 10, 100) & 100       & 120     & 150  & 30                            & 30                            \\ \hline
\end{tabular}
\end{table*}

Based on Figure \ref{Fig:topology}, a dumbbell topology was built. The link configuration is shown in Table \ref{Tab:dumbbell}, which only a small part of all the experiments.  These nodes are involved in this experiment are$n1, n2, n3, n5, n7, n8$. The shared bottleneck link is $L2$. There are three parameters $(B, D, Q)$ on every link, which mean link bandwidth (Mbps), link transmission delay (ms) and max tolerant queue delay (ms). The maximum queue delay is converted to maximum packets queue length in experiment according to link bandwidth, $q=\frac{BQ}{MTU}$. Once the number of queued packets exceed the configured queue length, the link is in state of congestion and the following incoming packets will be dropped. That’s the case for the Drop tail configuration, but in the RED queue situation, the bottleneck $L2$ was configure with RED queue management algorithm. The parameter sets of RED, minTh, maxTh, QL (maximum queue limit) are in unit of millisecond. When the queue length of the link is smaller than minTh$*B$, no packets will be dropped; when the queue length is between minTh$*B$ and maxTh$*B$, the incoming packets will be dropped according to some probability; all the incoming packets will be dropped once the queue length exceeds maxTh$*B$. The flows parameter, for example, flows (n1-n7), indicates that the total number of TCP congestions are initiated from node n1 to n7. On every topology experiments, all the connections are configured with the same congestion control algorithm Reno or Cubic. Thus, with different queue management mechanisms in bottleneck and two different congestion control algorithms of TCP connection, there are 20 experiments with the configuration in Table \ref{Tab:dumbbell}. Each experiment lasts 300 seconds.

\begin{figure}
\includegraphics[height=5in, width=3in]{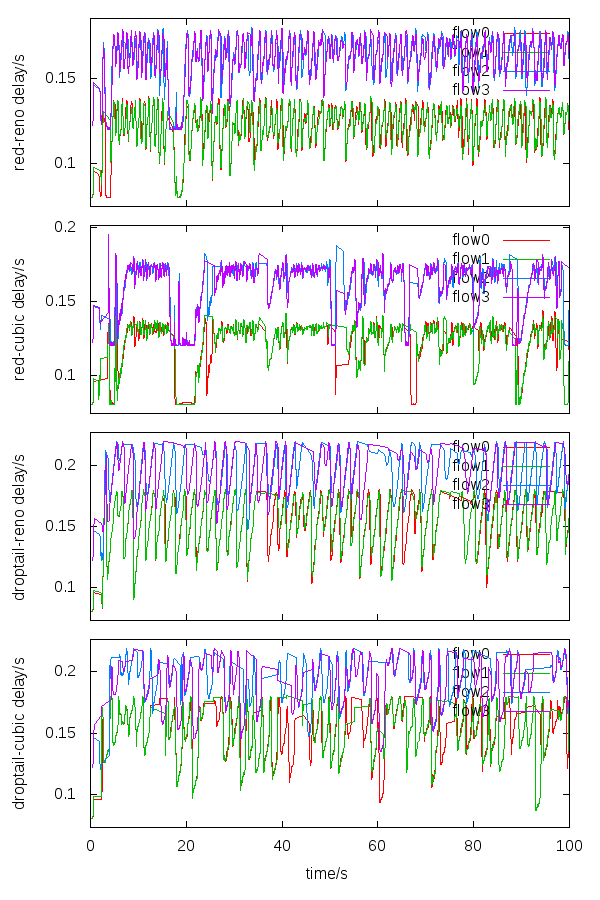}
\caption{The delay trace of tcp flows}
\label{Fig:exp5-delay}
\end{figure}

As an example, Figure \ref{Fig:exp5-delay} plots the changes of round trip delay of four connections (flow0 and flow1 from n1 to n7, flow2 and flow3 from n2 to n8) of the four experiments with the topology configuration of E5. Only these data samples from 0 to 100s are plotted on Figure \ref{Fig:exp5-delay}. The delay changes on these curves is the results of link queue filling by flows’ bandwidth probing phase and queue emptying due to flows’ rate reduction. The simulation duration of 300s on each experiment generates enough data samples and is enough to verify the effectiveness of shared bottleneck detection algorithms. From Figure \ref{Fig:exp5-delay}, the delay curves from connections through the same path show quite similarity and these curves from connections through totally different paths show some difference. But the slopes of different flows passing the same narrow link during the queue filling period are quite similar.

\begin{figure}
\includegraphics[height=5in, width=3in]{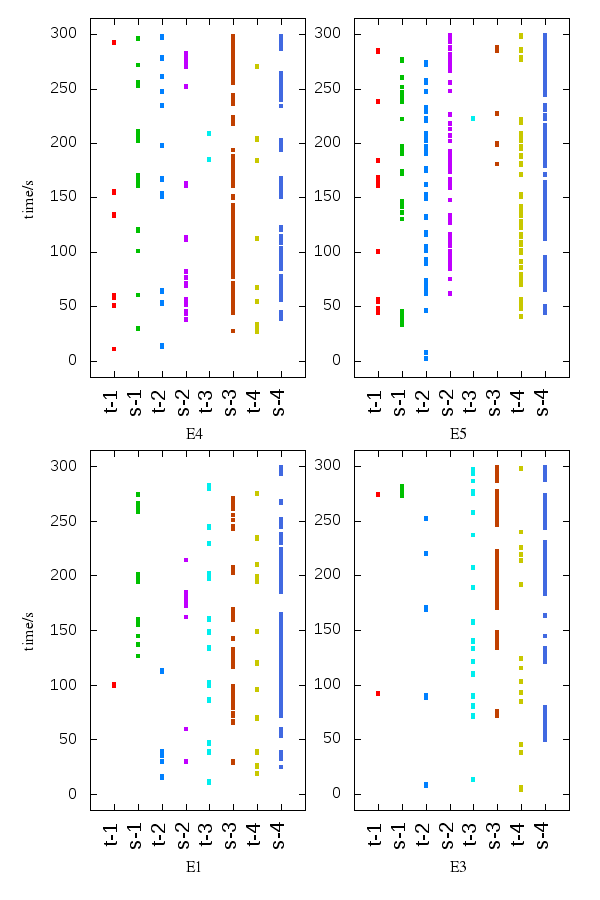}
\caption{The effective time interval detecting out shared bottleneck}
\label{Fig:sbd-compare}
\end{figure}

We verify our methods based on these collected data. The comparison with SSC is also conducted. The time intervals during which the TR and SSC methods give positive result indicating that two connections are via the same bottleneck are collected. And the final results are plotted in Figure \ref{Fig:sbd-compare}. The xtics t-1 or s-1, ``t`` denotes TR method and ``s`` denotes SSC method. The numbers (1, 2, 3, 4) after the dash line denote the combination of bottleneck queue configuration (Droptail or RED) and the congestion control algorithm (Cubic or Reno) of the TCP connections, which  (1, 2, 3, 4) denote Droptail-Cubic, Droptail-Reno, RED-Cubic and RED-Reno in respectively. 

Both TR and SSC work effectiveness in detecting out the shared bottleneck in most cases. And there is case that SSC fails for example the case label s-2 of E3 in Figure \ref{Fig:sbd-compare}. The effective time intervals of TR are sparse compared to SSC. Since TR method is only fed with data samples during the queue filling period and the SSC method is fed with all the traced data points. But our method reacts quickly to detect out shared bottleneck, 13/16 according to Figure \ref{Fig:sbd-compare}. That’s may be explained that the SSC method is based on statistical model. Only when enough collected data samples are fed to the SSC algorithm, can it find the similarity in data from two different connections. The ability to detect out shared bottleneck earlier can make the multipath protocol behave more fairness in coexisting with single path transmission protocol. In terms of computational complexity, according to the instruction in its paper, after every 350ms, the SSC runs with the data samples collected over the last 50 intervals while TR runs every five seconds. There's more computation overhead of SSC compared with TR. 

\subsection{Results from Multipath QUIC}
The shared bottleneck detection algorithm TR and SSC are implemented on multipath QUIC. And the performance of the two algorithm is compared on Mininet environment. First, the MPTCP LIA congestion control algorithm is transplanted on multipath QUIC. Based on LIA, the SBD algorithms TR and SSC are implemented. On every incoming acknowledgement packet, the timestamp when the acknowledge packet arrives and the smoothed round trip delay are fed to the SBD algorithms. At initialization, the none shared bottleneck is assumed, thus the endpoint configured with the uncoupled congestion control algorithm. Once the SBD algorithm gives positive signal, the endpoint will run LIA protocol. If the time interval exceeds an effective time threshold-which is 100s in current implementation, since the last time that detecting out shared bottleneck, the endpoint will fall back to uncouple congestion control protocol. That's the consideration of route dynamics in real network.

As mentioned before, in the shared bottleneck situation, the earlier SBD algorithm releases positive signal, the more fair the resource occupation of multipath transmission protocol in coexisting with single path TCP. A modified version of Jain's fairness index \cite{Jain1984Quantitative} is applied for fairness measurement. In equation \eqref{eq:fairness}, $\overline {x_{sp}}$ denotes the average rate of the random chosen TCP connections and $x_{mp}$ is the average rate multipath QUIC of all its sub-flows. The closer Jain's fairness index is to 1, the better in terms of resource fairness. In this experiment, the number of single TCP connections chosen for fairness computation is 10. The original Jain's fairness index equation exploits the rate of all the connections for fairness computation. During the test, only one multipath QUIC session was initiated. If the original Jain's fairness index is used, the final result will not be obvious and the impact of multipath QUIC on fairness computation can be neglected in account of the large number of TCP connections.
\begin{equation}
\label{eq:fairness}
J=\frac{(\overline {x_{sp}}+x_{mp})^2}{2*({\overline {x_{sp}}}^2+{x_{mp}}^2)}
\end{equation}

Two topologies are built in Mininet platform. Topology1 is consisted by nodes (n0, n1, n2, n3, n5) and links (L0, L1, L2, L7, L13). Topology2 is consisted of nodes (n0, n1, n2, n3, n5, n7, n8, n10, n11) and links (L0, L1, L2, L3, L13, L7, L10, L17, L18). The multi-homed QUIC client is started on n0, and the QUIC server is listening on n5 in topology1. The reason to build such topology is that the multipath QUIC session can only be initiated from client side in its golang codebase.

\begin{figure}
\includegraphics[height=2.5in, width=3in]{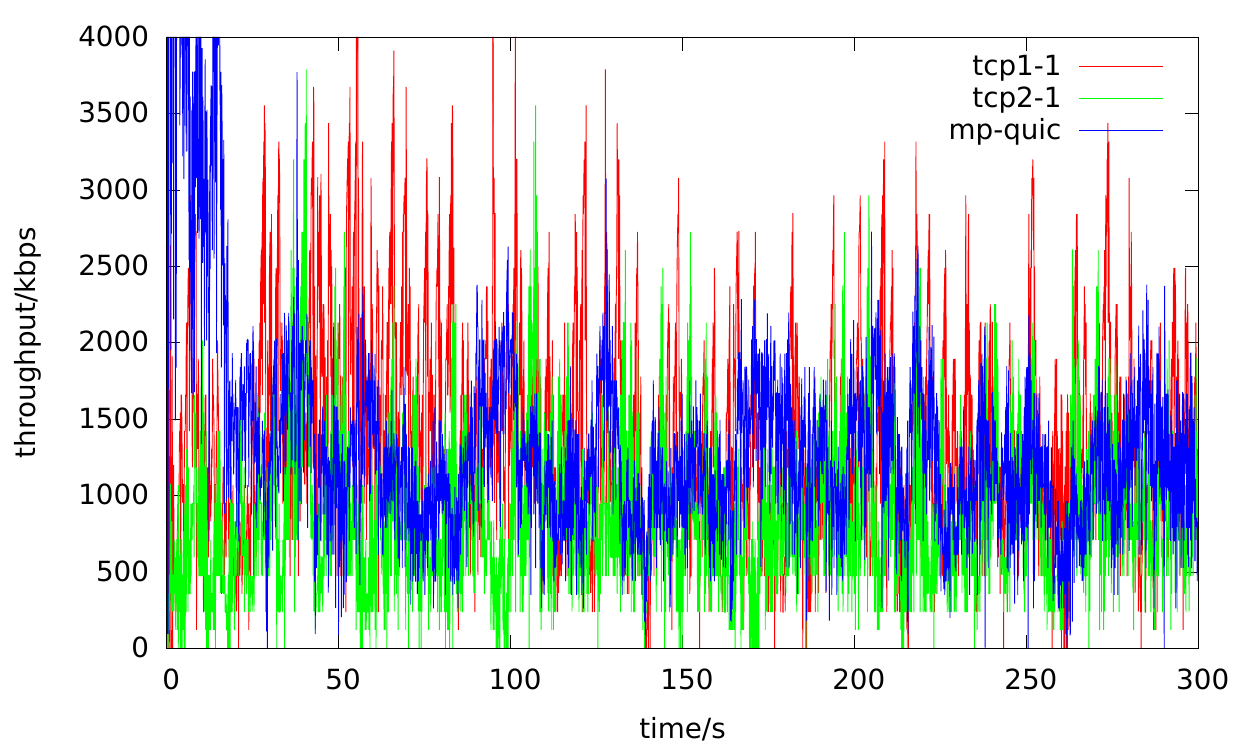}
\caption{The throughput of TR-LIA}
\label{Fig:tr-rate}
\end{figure}
\begin{figure}
\includegraphics[height=2.5in, width=3in]{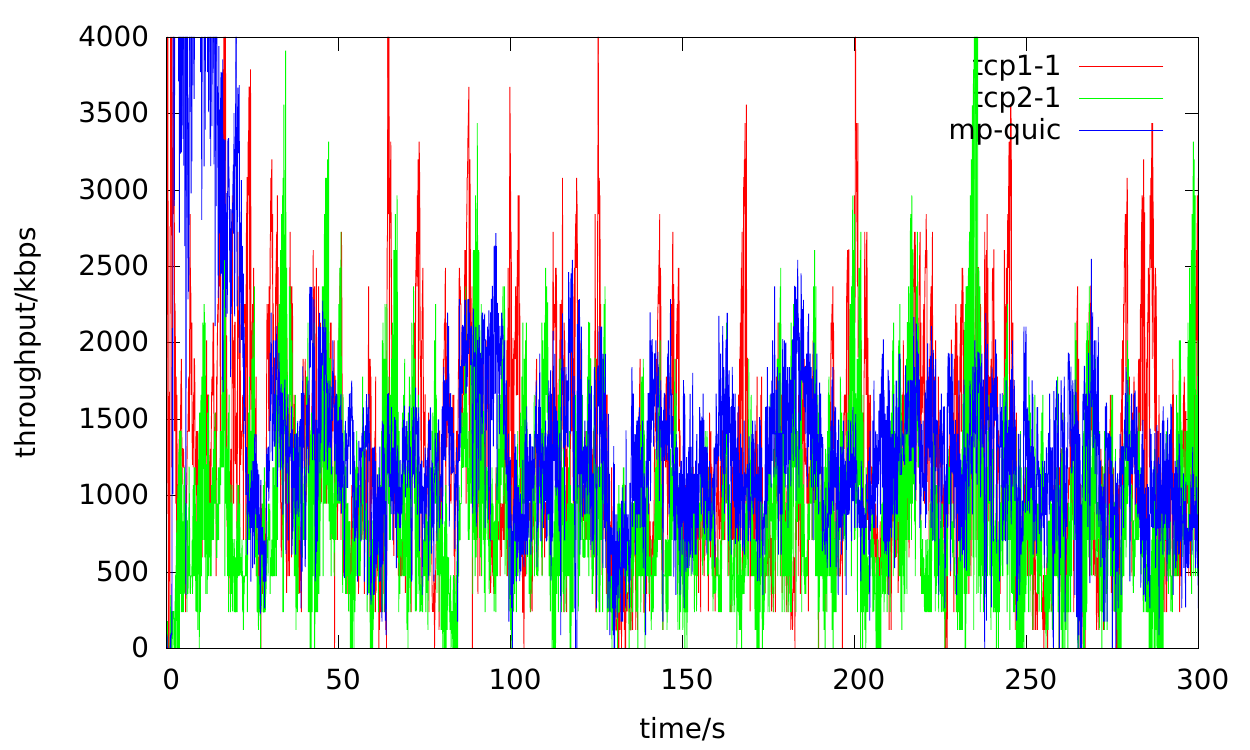}
\caption{The throughput of SSC-LIA}
\label{Fig:ssc-rate}
\end{figure}

\begin{table}[]
\caption{Topology1 link configuration}
\label{Tab:topology1}
\scalebox{0.6}{
\begin{tabular}{|c|c|c|c|c|c|c|c|c|c|c|}
\hline
   & L0            & L1            & L2         & L13           & L7            & Q    & f1  & f2 & $J_{SSC}$ & $J_{TR}$ \\ \hline
E1 & (500,10,10*Q) & (500,10,10*Q) & (200,10,Q) & (500,20,10*Q) & (500,20,10*Q) & 200  & 20  & 30 & 0.9952    & 0.9987   \\ \hline
E2 & (500,10,10*Q) & (500,10,10*Q) & (200,50,Q) & (500,20,10*Q) & (500,20,10*Q) & 200  & 50  & 50 & 0.9884    & 0.9993   \\ \hline
E3 & (500,10,10*Q) & (500,10,10*Q) & (300,40,Q) & (500,10,10*Q) & (500,20,10*Q) & 1000 & 60  & 70 & 0.9580    & 0.9677   \\ \hline
E4 & (800,10,10*Q) & (800,30,10*Q) & (400,20,Q) & (800,10,10*Q) & (800,20,10*Q) & 800  & 60 & 160 & 0.9479    & 0.9832   \\ \hline
\end{tabular}}
\end{table}

The link configuration of Topology1 is listed on Table \ref{Tab:topology1}. On every link, there are three tuples, which denotes bandwidth (Mbps), delay in millisecond and queue length in unit of packets. L2 is the bottleneck link. The parameter f1 is the number of TCP flows from n0 to n5 via n1 and f2 denotes TCP flows from n0 to n5 via n2. In every experiment, packets are captured with TCPDUMP on n5. The parameter J is the computed Jain’s fairness index. The multipath QUIC with LIA-TR and LIA-SSC are running separately on each topology configuration.

The route from n0 via n1 destined n5 is denoted as path1 and the route via n2 is denoted as path2. The throughput of two TCP flows coming from path1 and path2 and the rate of multipath QUIC session are computed on every 100 milliseconds. As an example, the rate results are shown in Figure \ref{Fig:tr-rate} and \ref{Fig:ssc-rate}, and tcp1-1 is a random chosen TCP flow from path1 whereas tcp2-1 is from path2. The average rates from these curves are quite close. From the two rates figures, the multipath QUIC endpoint configured with TR-LIA can restrict its packets sending rate more earlier compared with SSC due to the early shared bottleneck detection property.

\begin{table*}[]
\caption{Topology2 link configuration}
\label{Tab:topology2}
\scalebox{0.65}{
\begin{tabular}{|c|c|c|c|c|c|c|c|c|c|c|c|c|c|c|c|c|c|}
\hline
   & L0            & L1            & L2            & L3            & L11           & L13           & L7            & L10           & L17            & L18           & Q    & f0 & f1  & f2 & f3  & $J_{SSC}$ & $J_{TR}$ \\ \hline
E1 & (500,10,10*Q) & (500,10,10*Q) & (100,10,Q)    & (100,10,10*Q) & (100,10,10*Q) & (500,20,10*Q) & (500,10,10*Q) & (100,10,10*Q) & (100,30,10*Q)  & (100,10,10*Q) & 500  & 20 & 40  & 10 & 5   & 0.9410    & 0.9395   \\ \hline
E2 & (500,10,10*Q) & (500,10,10*Q) & (200,10,Q)    & (500,10,10*Q) & (500,10,10*Q) & (500,20,10*Q) & (500,10,10*Q) & (500,10,10*Q) & (500,30,10*Q)  & (500,10,10*Q) & 800  & 30 & 40  & 10 & 10  & 0.9355    & 0.9425   \\ \hline
E3 & (500,30,10*Q) & (500,10,10*Q) & (200,40,Q)    & (500,10,10*Q) & (500,10,10*Q) & (500,20,10*Q) & (500,10,10*Q) & (500,10,10*Q) & (500,30,10*Q)  & (500,10,10*Q) & 1500 & 30 & 40  & 10 & 10  & 0.8080    & 0.8078   \\ \hline
E4 & (500,30,10*Q) & (500,10,10*Q) & (400,40,Q)    & (500,10,10*Q) & (500,10,10*Q) & (500,20,10*Q) & (500,10,10*Q) & (500,10,10*Q) & (500,30,10*Q)  & (500,10,10*Q) & 1500 & 20 & 120 & 50 & 100 & 0.9726    & 0.9941   \\ \hline
E5 & (800,30,10*Q  & (800,10,10*Q) & (800,40,10*Q) & (800,10,10*Q) & (800,10,10*Q) & (800,20,10*Q) & (800,10,10*Q) & (800,10,10*Q) & (800,100,10*Q) & (300,10,Q)    & 1000 & 50 & 100 & 50 & 50  & 0.8953    & 0.9790   \\ \hline
\end{tabular}}
\end{table*}

As for Topology 2, more complex test cases can be configured. Table \ref{Tab:topology2} are chosen example during the experiments. The parameters f0, f1, f2, f3 denote the number of TCP connections from n0 via n1 to n11, n0 via n2 to n11, n1 to n7 and n7 to n8. As the Jain’s fairness index columns in Table \ref{Tab:topology1} and \ref{Tab:topology2} show, most experiment examples indicate the TR method is more fairness in bandwidth occupation. In case of E1 and E3 in Table \ref{Tab:topology2}, the value of $J_{TR}$ is junior to $J_{SSC}$, but they are quite close, since the first time that the two algorithms release positive signal are quite close. The slopes of two subflows in muitpath QUIC LIA are logged during the running of TR. These delay trend line slopes shown in Table \ref{Tab:slope} of E5 in Topology 2 are effective values that indicate the two flows sharing the same bottleneck.


In case of none-shared bottleneck configuration, the performance of TR and SSC is also compared. Because of the time gap between two congestion process, the effective slope value indicating the sharing bottleneck of sub-flows of TR is quite sparsity, which makes the TR more robust against false positive detection signal in none-SBD case compared to SSC and the endpoint can occupy more network bandwidth resource. Based on topology 1 and 2, the none-shared bottleneck link testing cases were configured and the common links will not be configured as the narrowest link. 

A throughput ratio is defined for algorithms performance comparison in equation \eqref{eq:ratio}. $\overline {x_{sp}}$ is the average throughput of random chosen single path TCP connections on the best path initialized from the same node with the multipath QUIC session. The difference in terms of throughput from Mininet platform is the reason to such definition. The throughput ratio comparison will be more meaningful and eliminates such uncertain bias. 
\begin{equation}
\label{eq:ratio}
R=\frac{x_{mp}}{\overline {x_{sp}}}
\end{equation}


\begin{table}[]
\caption{The effective slope in TR algorithm}
\label{Tab:slope}
\begin{tabular}{|c|c|c|}
\hline
time/s & slope1(ms/s)     & slope2(ms/s)     \\ \hline
9.874    & 473.00 & 441.00 \\ \hline
45.287   & 219.00 & 190.00 \\ \hline
50.523   & 397.00 & 454.00\\ \hline
70.634   & 140.00 & 147.00 \\ \hline
75.639   & 447.00 & 410.00 \\ \hline
121.007  & 348.00& 317.00 \\ \hline
136.115  & 335.00 & 341.00 \\ \hline
156.147  & 182.00 & 190.00 \\ \hline
166.215  & 286.00 & 283.00 \\ \hline
\end{tabular}
\end{table}
\section{Conclusion and future work}
In this paper, a delay trend line regression method is proposed for sharing bottleneck detection in multipath transmission protocol. Deduced from TCP fluid model, these flows traversing a common bottleneck shows linear increase in packet round trip delay signal during the link queue building up process. The delay slopes calculated by linear regression of two flows under a predefined threshold are considered traversing a common bottleneck. In order to extract out these data samples during the link queue filling process, a max value comparison filter is applied. The effectiveness of the proposed algorithm is validated based on offline data analysis traced from ns3 platform, in topology configured with RED or Droptail queue management mechanism and TCP sessions with Reno or cubic congestion control algorithm.

Further, the TR is incorporated in LIA on multipath QUIC codebase and its performance is compared with SSC on Mininet platform in terms of fairness in shared bottleneck configuration and throughput improvement in none shared bottleneck case. Results shows that TR is more fairness when completing resource with single TCP connections in case of the two subflows of the multipath session transporting the same bottleneck, which is beneficial from its earlier detection feature. As for the none sharing bottleneck, both TR and SSC show detection error in some cases. Once a false positive signal is generated, the multipath session would fall back to coupled congestion control in the following 100 second in implementation, which has a negative impact on throughput. TR algorithm gets faster recovery from false positive signal and tends to occupy more bandwidth in comparison with SSC method due to the sparse queue delay signal. The accuracy comparison of the two algorithm cannot be provided. As shown in \cite{Ferlin2016Revisiting}, the mean accuracy of SSC is about 70\%. If the SBD algorithm releases false positive results, the congestion control of multipath protocol will fall back to its original coupled form and there is no obvious harm. The merit overweighs its flaws since the bandwidth improvement of the multi-homed endpoint in none sharing bottleneck so obvious that makes the combination of multipath congestion control and SBD mechanism an attractive feature.

Both TR and SSC are based on passive data collection without extra packets injecting into the networks, which may be the reason of its accuracy loss in none sharing bottleneck situation. The combination of active probing and signal processing may be one possible solution for accuracy improvement which will remain our future work.

The newly proposed congestion control mechanism BBR has achieved state of art improvement for TCP, which makes the research on congestion control a hotspot once again. And we found in experiment, the multipath protocol nearly starves to death when completing with TCP BBR. As the BBR has been implemented in Linux kernel 4.10 and later version, which could be widely applied in short future. One of the goal of the MPTCP is throughput improvement to motivate deployment is in challenge. Even through there is advice to implement BBR on multipath TCP, such implementation could obviously result unfairness in network bandwidth allocation. And an effective sharing bottleneck detection algorithm may work as a solution to meet the requirement the throughput improvement and bandwidth fairness when its subflows sharing the same bottleneck, if the BBR congestion control algorithm is applied in multipath transmission protocol. That’s could be another direction of our future work.
\section{acknowledgments }
This work was supported by the National Natural Science Foundation of China [No. 61671141, No. 61401081],  the Liaoning Provincial Natural Science Foundation of China [No. 20180551007], and the Ministry of Education-China Mobile Scientific Research Funds [No. MCM20150103].

\bibliographystyle{unsrt}
\bibliography{sbd}

\begin{thebibliography}{10}

\bibitem{Holterbach2017SWIFT}
Thomas Holterbach, Stefano Vissicchio, Alberto Dainotti, and Laurent Vanbever.
\newblock Swift: Predictive fast reroute.
\newblock In {\em Proceedings of the Conference of the ACM Special Interest
  Group on Data Communication}, SIGCOMM '17, pages 460--473, New York, NY, USA,
  2017. ACM.

\bibitem{Iyengar2006Concurrent}
J.~R. Iyengar, P.~D. Amer, and R.~Stewart.
\newblock Concurrent multipath transfer using sctp multihoming over independent
  end-to-end paths.
\newblock {\em IEEE/ACM Transactions on Networking}, 14(5):951--964, Oct 2006.

\bibitem{RFC6182}
Alan Ford, Costin Raiciu, Mark Handley, Sebastien Barre, and Janardhan Iyengar.
\newblock Architectural guidelines for multipath tcp development, Mar 2011.
\newblock RFC6182.

\bibitem{rfc6356}
Costin Raiciu, Mark Handley, and Damon Wischik.
\newblock Coupled congestion control for multipath transport protocols.
\newblock RFC 6356, RFC Editor, Oct 2011.

\bibitem{Cao2012Delay}
Yu~Cao, Mingwei Xu, and Xiaoming Fu.
\newblock Delay-based congestion control for multipath tcp.
\newblock In {\em 2012 20th IEEE International Conference on Network Protocols
  (ICNP)}, pages 1--10, Oct 2012.

\bibitem{Khalili2013MPTCP}
Ramin Khalili, Nicolas Gast, Miroslav Popovic, and Jean-Yves Le~Boudec.
\newblock Mptcp is not pareto-optimal: Performance issues and a possible
  solution.
\newblock {\em IEEE/ACM Trans. Netw.}, 21(5):1651--1665, October 2013.

\bibitem{peng2016multipath}
Q.~Peng, A.~Walid, J.~Hwang, and S.~H. Low.
\newblock Multipath tcp: Analysis, design, and implementation.
\newblock {\em IEEE/ACM Transactions on Networking}, 24(1):596--609, Feb 2016.

\bibitem{Cardwell2017BBR}
Neal Cardwell, Yuchung Cheng, C.~Stephen Gunn, Soheil~Hassas Yeganeh, and Van
  Jacobson.
\newblock Bbr: Congestion-based congestion control.
\newblock {\em Commun. ACM}, 60(2):58--66, January 2017.

\bibitem{jacobson1988congestion}
V.~Jacobson.
\newblock Congestion avoidance and control.
\newblock {\em SIGCOMM Comput. Commun. Rev.}, 18(4):314--329, August 1988.

\bibitem{Nichols2012Controlling}
Kathleen Nichols and Van Jacobson.
\newblock Controlling queue delay.
\newblock {\em Communications of the ACM}, 55(7):42--50, 2012.

\bibitem{Balakrishnan1999Integrated}
Hari Balakrishnan, Hariharan~S. Rahul, and Srinivasan Seshan.
\newblock An integrated congestion management architecture for internet hosts.
\newblock {\em SIGCOMM Comput. Commun. Rev.}, 29(4):175--187, August 1999.

\bibitem{Andersen2002Resilient}
David Andersen, Hari Balakrishnan, Frans Kaashoek, and Robert Morris.
\newblock Resilient overlay networks.
\newblock {\em SIGCOMM Comput. Commun. Rev.}, 32(1):66--66, January 2002.

\bibitem{Zhang2014general}
Wei Zhang, Weimin Lei, Shaowei Liu, and Guangye Li.
\newblock A general framework of multipath transport system based on
  application-level relay.
\newblock {\em Computer Communications}, 51:70 -- 80, 2014.

\bibitem{Guan2018Scalable}
Yunchong Guan, Weimin Lei, Wei Zhang, Shaowei Liu, and Hao Li.
\newblock Scalable orchestration of software defined service overlay network
  for multipath transmission.
\newblock {\em Computer Networks}, 137:132 -- 146, 2018.

\bibitem{Katabi2001passive}
D.~Katabi, I.~Bazzi, and Xiaowei Yang.
\newblock A passive approach for detecting shared bottlenecks.
\newblock In {\em Proceedings Tenth International Conference on Computer
  Communications and Networks (Cat. No.01EX495)}, pages 174--181, 2001.

\bibitem{Rubenstein2002Detecting}
D.~Rubenstein, J.~Kurose, and D.~Towsley.
\newblock Detecting shared congestion of flows via end-to-end measurement.
\newblock {\em IEEE/ACM Transactions on Networking}, 10(3):381--395, Jun 2002.

\bibitem{Younis2005FlowMate}
O.~Younis and S.~Fahmy.
\newblock Flowmate: scalable on-line flow clustering.
\newblock {\em IEEE/ACM Transactions on Networking}, 13(2):288--301, April
  2005.

\bibitem{Cui2004SCONE}
Weidong Cui, Sridhar Machiraju, Randy~H Katz, and Ion Stoica.
\newblock Scone: A tool to estimate shared congestion among internet paths.
\newblock {\em EECS Department, University of California, Berkeley, Tech. Rep.
  UCB/CSD-04-1320}, 2004.

\bibitem{Karacali2008Network}
B.~Karacali and M.~Karol.
\newblock Network-wide inference of end-to-end path intersections.
\newblock In {\em NOMS 2008 - 2008 IEEE Network Operations and Management
  Symposium}, pages 168--175, April 2008.

\bibitem{Wang2004Passive}
Lili Wang, James~N. Griffioen, Kenneth~L. Calvert, and Sherlia Shi.
\newblock Passive inference of path correlation.
\newblock In {\em Proceedings of the 14th International Workshop on Network and
  Operating Systems Support for Digital Audio and Video}, NOSSDAV '04, pages
  36--41, New York, NY, USA, 2004. ACM.

\bibitem{Zhu2013TCP}
W.~Zhu.
\newblock Tcp path sharing detection.
\newblock In {\em 2013 IEEE 11th Malaysia International Conference on
  Communications (MICC)}, pages 244--249, Nov 2013.

\bibitem{Kim2008Wavelet}
M.~S. Kim, T.~Kim, Y.~J. Shin, S.~S. Lam, and E.~J. Powers.
\newblock A wavelet-based approach to detect shared congestion.
\newblock {\em IEEE/ACM Transactions on Networking}, 16(4):763--776, Aug 2008.

\bibitem{Hayes2014Practical}
D.~A. Hayes, S.~Ferlin, and M.~Welzl.
\newblock Practical passive shared bottleneck detection using shape summary
  statistics.
\newblock In {\em 39th Annual IEEE Conference on Local Computer Networks},
  pages 150--158, Sept 2014.

\bibitem{Ferlin2016Revisiting}
S.~Ferlin, Ö. Alay, T.~Dreibholz, D.~A. Hayes, and M.~Welzl.
\newblock Revisiting congestion control for multipath tcp with shared
  bottleneck detection.
\newblock In {\em IEEE INFOCOM 2016 - The 35th Annual IEEE International
  Conference on Computer Communications}, pages 1--9, April 2016.

\bibitem{Hong2003Note}
Dohy Hong.
\newblock {A Note on the TCP Fluid Model}.
\newblock Research Report RR-4703, {INRIA}, 2003.

\bibitem{Chiu1989Analysis}
Dah-Ming Chiu and Raj Jain.
\newblock Analysis of the increase/decrease algorithms for congestion avoidance
  in computer networks. j-comp-net-isdn, 17 (1): 1--14, 1989.

\bibitem{Ha2008CUBIC}
Sangtae Ha, Injong Rhee, and Lisong Xu.
\newblock Cubic: A new tcp-friendly high-speed tcp variant.
\newblock {\em SIGOPS Oper. Syst. Rev.}, 42(5):64--74, July 2008.

\bibitem{ns3}
ns3.
\newblock \url{https://www.nsnam.org/}.

\bibitem{wangpan}
Baidu wangpan.
\newblock \url{https://pan.baidu.com/}.

\bibitem{DeConinck2017Multipath}
Quentin De~Coninck and Olivier Bonaventure.
\newblock Multipath quic: Design and evaluation.
\newblock In {\em Proceedings of the 13th International Conference on Emerging
  Networking EXperiments and Technologies}, CoNEXT '17, pages 160--166, New
  York, NY, USA, 2017. ACM.

\bibitem{Lantz2010Network}
Bob Lantz, Brandon Heller, and Nick McKeown.
\newblock A network in a laptop: Rapid prototyping for software-defined
  networks.
\newblock In {\em Proceedings of the 9th ACM SIGCOMM Workshop on Hot Topics in
  Networks}, Hotnets-IX, pages 19:1--19:6, New York, NY, USA, 2010. ACM.

\bibitem{Jain1984Quantitative}
Rajendra~K Jain, Dah-Ming~W Chiu, and William~R Hawe.
\newblock A quantitative measure of fairness and discrimination.
\newblock {\em Eastern Research Laboratory, Digital Equipment Corporation,
  Hudson, MA}, 1984.

\end{thebibliography}

\end{document}